\documentclass[jcp,twocolumn]{revtex4}
\usepackage{graphicx}
\usepackage{dcolumn}
\usepackage{amsmath}
\usepackage{amssymb}
\usepackage{rotating}
\usepackage{soul}  
\sethlcolor{yellow}
\usepackage{url}
\usepackage{hyperref}
\usepackage[dvips]{color}
\usepackage{color}
\definecolor{bluegreen}{rgb}{0,0.2,0.8}
\usepackage{subfigure,amsmath,verbatim,moreverb}

\def\Iv{{\bf I}}
\def\Jv{{\bf J}}
\def\Kv{{\bf K}}
\def\nn{\nonumber}
\definecolor{bluegreen}{rgb}{0,0.2,0.8}
\begin{document}
\title{Origin of the size-dependence of the equilibrium van der Waals binding
between nanostructures}
\author{Jianmin Tao}
\altaffiliation{Corresponding author: jianmin.tao@temple.edu \\
URL: \url{http://www.sas.upenn.edu/~jianmint/}}
\affiliation{Department of Physics, Temple University, Philadelphia, 
PA 19122-1801, USA}
\author{John P. Perdew}
\affiliation{Department of Physics, Temple University, Philadelphia,
PA 19122-1801, USA}
\author{Hong Tang}
\affiliation{Department of Physics, Temple University, Philadelphia,
PA 19122-1801, USA}
\author{Chandra Shahi}
\affiliation{Department of Physics, Temple University, Philadelphia,
PA 19122-1801, USA}

\date{\today}
\begin{abstract}
Nanostructures can be bound together at equilibrium by the van der Waals 
(vdW) effect, a small but ubiquitous many-body attraction that presents 
challenges to density functional theory. How does the binding energy 
depend upon the size or number of atoms in one of a pair of identical 
nanostructures? To answer this question, we treat each nanostructure 
properly as a whole object, not as a collection of atoms. Our calculations 
start from an accurate static dipole polarizability for each considered 
nanostructure, and an accurate equilibrium center-to-center distance for 
the pair (the latter from experiment, or from the vdW-DF-cx functional). 
We consider the competition in each term $-C_{2k}/d^{2k}$ ($k=3, 4, 5$) of 
the long-range vdW series for the interaction energy, between the size 
dependence of the vdW coefficient $C_{2k}$ and that of the $2k$-th power 
of the center-to-center distance $d$. The damping of these vdW terms can be 
negligible, but in any case it does not affect the size dependence for a 
given term in the absence of non-vdW binding. To our surprise, the vdW 
energy can be size-independent for quasi-spherical nanoclusters bound to 
one another by vdW interaction, even with strong nonadditivity of the vdW 
coefficient, as demonstrated for fullerenes. We also show that, for 
low-dimensional systems, the vdW interaction yields the strongest 
size-dependence, in stark contrast to that of fullerenes. We illustrate 
this with parallel planar polycyclic aromatic hydrocarbons. Other 
cases are between, as shown by sodium clusters.
\end{abstract}

\maketitle
\section{Introduction}
Conventional Kohn-Sham density functional theory (DFT) has
reached a high level of sophistication and achieved practical success, due 
to the good balance between efficiency and achievable accuracy. In recent 
years, many reliable semilocal density functionals have been 
proposed~\cite{pbe96,tpss03,M06L,SCAN,JSun16,TM16,excitation,pbe0,Becke1996} 
and some of them have been widely used in electronic structure calculations. 
However, these conventionally constructed DFT methods often produce large 
errors for molecular complexes and solids~\cite{peng16}, interface 
problems~\cite{tao-rappe14}, and ionic solids~\cite{tao17}. A fundamental 
reason is that, while conventional DFT methods can give an accurate 
description of the short-range part, the long-range vdW interaction is 
missing in these methods.

The long-range vdW interaction is an important nonlocal correlation due to
instantaneous electric charge fluctuations. It affects many properties
of molecular complexes and solids~\cite{Egger14,WAASaidi12,OHod12,ERJ12,
MScheffler12,Grimme13,book14,ERJ15,TYR15}, layered materials~\cite{BSachs11,
Anyan13,peng16}, and ionic solids~\cite{TGould16,tao17}. It has been shown 
that performances of conventional DFT methods can be substantially improved 
with a vdW correction. A number of accurate vdW corrections have been 
developed. Most of them are based on atom pairwise effective 
models~\cite{MScheffler09,SGrimme10,Becke07,Steinmann11-1,
Steinmann11-2}, while a few of them~\cite{SGrimme10,MDion04,KBerland14-1,review15,
Voorhis09,Granatier2011,JKlimes11} go beyond atom pairwise models. These 
models are very accurate for small or mid-size intermolecular interaction 
and have been widely used in electronic structure calculations, but errors 
may grow with system size or number of atoms in a system and can seriously 
affect the performances of vdW-corrected methods, as system size approaches 
the nanoscale. For example, it has been shown that, while the 
dispersion-corrected atom pairwise model PBE+D2 is accurate for the binding 
of a pair of small molecules, the errors become large for 
fullerenes~\cite{PRL12,TYR15}. It is known that an important source of 
errors is the nonadditivity of the vdW coefficients~\cite{DWR}, due to 
electron delocalization and many-body (e.g., three-body) 
interactions~\cite{Klein67,OAL10}. 
         
Additivity means that the multipole polarizability of a nanostructure scales 
linearly with $N$,the number of atoms in it, and that each vdW coefficient 
$C_{2k}$ between identical nanostructures scales as $N^2$, the behavior 
predicted by atom pairwise interactions. As we will see, this expectation is 
rarely exact even for the dipole polarizability and for $C_6$, and never for 
higher-order contributions. A proper treatment of non-additivity requires 
treating each nanostructure as a whole object, and not as a collection of 
atoms.

In solids, a particle at $\Iv$ not only interacts with another particle 
at $\Jv$, but also with pairs $\Jv\Kv$ of other particles, etc. The 
energy of vdW interaction of a particle at $\Iv$ with all others can be 
written as a many-body expansion~\cite{Bell66,Tang69,MBDoran71,OAL10,
Paesani16}
\begin{eqnarray}\label{three}
E_{\rm vdW}(\Iv) =  \sum_{\Jv}E_{\rm vdW}^{(2)}(\Iv\Jv)+
\sum_{\Jv\Kv}E_{\rm vdW}^{(3)}(\Iv\Jv\Kv)+\cdots
\end{eqnarray}
where $E_{\rm vdW}^{(2)}$ is the two-body contribution, while
$E_{\rm vdW}^{(3)}$ accounts for the three-body contribution.

\section{Two-body vdW interaction energy} 
A model for the two-body vdW interaction energy is usually developed from the 
asymptotic expansion of the vdW interaction at large separation, which 
can be written as
\begin{eqnarray}\label{vdw}
E_{\rm vdW}^{(2)} &=&
-(C_6/d^6)f_{d,6}(d/d_{\rm vdW})-(C_8/d^8)f_{d,8}(d/d_{\rm vdW}) 
\nn \\ &-&
(C_{10}/d^{10})f_{d,10}(d/d_{\rm vdW}),
\end{eqnarray}
where $d$ is the distance between the centers of two interacting density 
fragments and $f_{d,2k}$ is the damping function~\cite{Chai00, 
KTTang84,wuyang02,ERJ06,SGrimme11}, with $d_{\rm vdW}$ being the sum of 
vdW radii. In the development of vdW corrections, there are two important 
tasks. One is the calculation of vdW coefficients, and the other is the 
design of a proper damping function, according to the short-range 
interaction from a semilocal DFT. The former involves important many-body 
effects and has received most attention. It was shown~\cite{TYR15} that in 
nanostructured materials such as fullerenes, 
$f_{d,2k}(d/d_{\rm vdW}\sim 1) \approx 1$. 
The sum of vdW radii can define the minimum separation between the centers of 
density fragments for nanostructured materials without formation of a covalent 
bond. The center-to-center distance can be written as $d=d_{\rm vdW}+\Delta$, 
with $\Delta$ being determined by the nature of the interaction. It can
be positive, zero (i.e., direct contact), or negative. For nanostructures 
bound at equilibrium by only the vdW interaction, $d/d_{vdW}$ is of order unity, 
and nearly independent of system size (as we will show), so even when damping 
is important for the interaction energy it is not important for the 
size-dependence of each term in Eq.~(\ref{vdw}).

The two-body vdW coefficients can be calculated from the dynamic multipole 
polarizability via the Casimir-Polder formula~\cite{SHPatil97},
$C_{2k}^{\rm AB}=[(2k-2)!/(2\pi)]\sum_{l_1=1}^{k-2}
[1/(2l_1)!(2l_2)!]
\int_0^\infty du~\alpha_{l_1}^{\rm A}(iu)\alpha_{l_2}^{\rm B}(iu)$,
where $l_2=k-l_1-1$. The required dynamic multipole polarizability can be 
modelled with the spherical-shell model. Since the electron 
density in nanostructures is nearly independent of system size or number 
of atoms in a system, we may simplify the spherical-shell model with the 
single-frequency approximation (SFA)~\cite{JTao14,Tao-Rappe16}, in which 
we assume that (i) only valence electrons in the outermost subshell of an 
atom in a molecule are polarizable, and (ii) the density is uniform inside 
the effective or vdW radius $R_l$ and zero otherwise. This model is 
particularly useful for nanostructures or larger systems, in which the 
electron density is slowly-varying~\cite{JTao14,Tao-Rappe16,PRB16}. The 
only required input is the static polarizability which can be obtained 
from accurate {\em ab initio} methods. 

Within the SFA, the model dynamic multipole polarizability takes the 
simple expression
\begin{eqnarray}\label{sfapolar}
\alpha_{l}^{\rm SFA}(iu) = R_l^{2l+1}\frac{\omega_l^2}{\omega_l^2+u^2}
~\frac{1-\rho_l}{1-\beta_l\rho_l},
\end{eqnarray}
where $R_l$ is the effective outer radius of the shell defined below 
[Eq.~(\ref{conventionalP})], $\beta_l = \omega_l^2~ {\tilde{\omega}}_l^2/
[(\omega_l^2+u^2)({\tilde{\omega}}_l^2+u^2)]$ describes the coupling of
the sphere and cavity plasmon oscillations, and $\rho_l = 
(1-t_l/R_l)^{2l+1}$ describes the shape of the shell, with $t_l$ being the 
shell thickness~\cite{GKG04,JTao14}. In the static limit, the model dynamic 
multipole polarizability reduces to the true static polarizability, i.e., 
$\alpha_{l}^{\rm SFA}(0) = \alpha_{l}(0)$. For fullerenes, we set 
$t_l=3.4$ bohr. $\omega_l = \omega_p \sqrt{l/(2l+1)}$ is the average sphere 
plasmon frequency, ${\tilde{\omega}}_l = \omega_p\sqrt{(l+1)/(2l+1)}$ is 
the cavity plasmon frequency, and $\omega_p = \sqrt{4\pi {\bar n}}$ is the 
average plasmon frequency of the extended electron gas, with ${\bar n} = 
{\cal N}/V_l$ and $V_l$ being the effective vdW volume and $\cal N$ the 
total number of valence electrons in the outermost subshell (${\cal N}=2$ 
for carbon atom, while for B and N atoms, ${\cal N}=1$, $3$).

For a classical conducting shell of uniform density, the static multipole 
polarizability is related to the static dipole polarizability by
$\alpha_l(0) = [\alpha_1(0)]^{(2l+1)/3}$. The vdW radius is defined by
\begin{eqnarray}\label{conventionalP}
R_l=[\alpha_l(0)]^{1/(2l+1)},
\end{eqnarray}
with $R_1=R_2=R_3 =[\alpha_1(0)]^{1/3}$. The preceding formula, which 
predicts the static higher-order polarizabilities from the static dipole
polarizability, is valid for slowly-varying densities, but only 
approximately true for strongly inhomogeneous densities such as atoms and 
ions~\cite{tao17}. It can be shown that the average electron density 
${\bar n}$ within the shell is nearly a constant for nanostructures such 
as fullerenes. Therefore, we can write
\begin{eqnarray}\label{vdw1}
E_{\rm vdW}^{(2)}&=&-
\frac{\alpha_1^A(0)\alpha_1^B(0)f_{11}({\bar n_A},{\bar n_B})}
{([\alpha_1^A(0)]^{1/3}+[\alpha_1^B(0)]^{1/3}+\Delta)^6}
\nn \\ &-&
\frac{\alpha_1^A(0)\alpha_2^B(0)f_{12}({\bar n_A},{\bar n_B})+P_{21}}
{([\alpha_1^A(0)]^{1/3}+[\alpha_1^B(0)]^{1/3}+\Delta)^8}
\nn \\ &-&
\frac{\alpha_1^A(0)\alpha_3^B(0)f_{13}({\bar n_A},{\bar n_B})+P_{22}+P_{31}}
{([\alpha_1^A(0)]^{1/3}+[\alpha_1^B(0)]^{1/3}+\Delta)^{10}},
\end{eqnarray}
where $f_{l_1l_2}({\bar n_A},{\bar n_B})$ represents the integral over 
the imaginary frequency 
$iu$, whose explicit expression can be extracted from Ref.~\cite{PRL12}. 
$P_{21}$, $P_{22}$ and $P_{31}$ represent the terms containing
$\alpha_2^A(0)\alpha_1^B(0)$, $\alpha_2^A(0)\alpha_2^B(0)$,
and $\alpha_3^A(0)\alpha_1^B(0)$, respectively.
Note that Eq.~(\ref{vdw1}) is valid for any two nanotructures, 
no matter whether they are identical or not. Consider a sequence of systems
in which system size or number of atoms in a system increases from the 
initial ($i$) to final ($f$) size. The static multipole polarizability also 
changes from $\alpha_{l}^i(0)$ to $\alpha_{l}^f(0)=
{\cal N}_f^{1+\delta_l}[\alpha_{l}^i(0)/N_i]$ for ${\cal N}_f \gg {\cal N}_i$. 
Here $\delta_l$ is the nonadditivity of the static multipole polarizability 
measuring the deviation of the conventional value $\alpha_{l}^{conv,f}(0)
={\cal N}_f[\alpha_{l}^i(0)/{\cal N}_i]$ from the true value 
$\alpha_{l}^f(0)$. For example, $\delta_1=0.2$ for the dipole polarizability
of fullerenes~\cite{JCP13,NatureM}, and $-0.084$ for sodium clusters evaluated
from the dipole polarizability of Ref.~\cite{AJiemchooroj06}. As seen below, 
even when $\delta_1 = 0$, $\delta_l > 0$ for $l \ge 2$.

\begin{table*}
\caption{Terms of the van der Waals interaction energy for a pair of 
identical fullerenes in a fullerene solid 
~\cite{TYR15} with
the nearest neighbor center-to-center separation $d_{cc}$ obtained from fcc
experimental lattice constants~\cite{MNMagomedov05,RDBendale95}, and 
for a pair of identical sodium clusters. The dipole 
polarizabilities of Na$_{4}$ and Na$_{8}$ are from 
Ref.~\cite{AJiemchooroj06}, while that of Na$_{19}$ is obtained as 
$\alpha_1(0)=(R+\delta)^3$, with $R=N^{1/3}r_s$ and $\delta=1.5$ 
bohr~\cite{DEBeck84}. $d_{cc}$ is calculated as the distance between 
the centers of mass of two sodium clusters by putting them side by 
side~\cite{PRA02} (AA stacking for $({\rm Na}_4)_2$) 
with Quantum ESPRESSO~\cite{QE} using the vdW-DF-cx 
functional~\cite{KBerland14-1}. All quantities are in atomic units. 
The reference values for $C_6$ are from Ref~\cite{JCP13} for fullerenes 
and Ref.~\cite{AJiemchooroj06} for sodium clusters, except for Na$_{19}$, 
which is taken as an average of Na$_{18}$ and Na$_{20}$. The calculated 
vdW coefficients~\cite{JTao14,PRB16} are obtained from the hollow-sphere 
model within SFA of Eq.~(\ref{sfapolar}). A minus sign ``$-$'' in front of 
all the vdW interactions has been suppressed.}
\small
\begin{tabular}{l|cccccccccccccc}
\hline\hline
&
\multicolumn{1}{c}{$\alpha_1$}&
\multicolumn{1}{c}{$C_6^{\rm ref}/10^3$}&
\multicolumn{1}{c}{$C_6/10^3$}&
\multicolumn{1}{c}{$C_8/10^5$}&
\multicolumn{1}{c}{$C_{10}/10^8$}&
\multicolumn{1}{c}{$C_9/10^5$}&
\multicolumn{1}{c}{$d_{cc}$}&
\multicolumn{1}{c}{$d_{\rm vdW}$}&
\multicolumn{1}{c}{$\frac{d_{cc}}{d_{\rm vdW}}$}&
\multicolumn{1}{c}{$\Delta$}&
\multicolumn{1}{c}{$10^3C_6/d^6$}&
\multicolumn{1}{c}{$10^3C_8/d^8$}&
\multicolumn{1}{c}{$10^3C_{10}/d^{10}$}&
\multicolumn{1}{c}{$(C_9/d^9) 10^3$} \\ \hline
C$_{60}$-C$_{60}$ &536.6&100.1&98.91&356.9&105.9&396.0&18.9&16.3&1.16&2.6 &2.2&2.4&2.0&0.13 \\
C$_{70}$-C$_{70}$ &659.1&141.6&144.7&601.8&205.7&711.1&20.1&17.4&1.16&2.7 &2.2&2.5&2.2&0.13 \\
C$_{78}$-C$_{78}$ &748.3&178.2&184.2&836.1&311.9&1027 &20.8&18.2&1.14&2.6 &2.2&2.5&2.2&0.14 \\
C$_{84}$-C$_{84}$ &806.1&207.7&213.3&1019 &400.2&1281 &21.5&18.6&1.16&2.9 &2.2&2.5&2.2&0.13 \\
\hline
Na$_{4}$-Na$_{4}$ &511.5&16.80&17.28&51.40&13.75&57.89&13.0&16.0&0.81&-2.7&3.6 &6.3 &10 &0.55 \\
Na$_{8}$-Na$_{8}$ &883.9&52.48&55.68&251.3&98.47&342.5&15.7&19.2&0.82&-3.3&3.7 &6.8 &11 &0.59 \\
Na$_{19}$-Na$_{19}$&1804&241.9&250.5&1941 &1249 &3389 &16.5&24.3&0.68&-7.8&12.4&35.3&84 &3.74 \\
\hline\hline
\end{tabular}
\label{table1}
\end{table*}

In the preceding paragraph,
$\delta_l$ is a measure of nonadditivity of the static multipole 
polarizability, as explained above. It is given by~\cite{NatureM}
$\delta_l= 
[(2l+1)(1+\delta_1)-3]/3-(1/3)[(2l+1)-3]({\rm ln} 
{\cal N}_i/{\rm ln} {\cal N}_f)$,
where the second term is a correction to the first term. 
For ${\cal N}_i=1$ or ${\cal N}_i \ll {\cal N}_f$, the second term
vanishes. Then, we obtain $\delta_2=(2+5\delta_1)/3$
and $\delta_3=(4+7\delta_1)/3$. Substituting the expression for the 
nonadditivity of the static multipole polarizability into Eq.~(\ref{vdw1}) 
and considering A = B leads to a dramatically simplified 
size-dependence of the vdW interaction energy between nanostructures,
\begin{eqnarray}\label{vdw2}
E_{\rm vdW}^{(2)}&=&-
\frac{[\alpha_{1}^i(0)]^2f_1(\bar n)}
{(2[\alpha_{1}^i(0)]^{1/3}+\Delta/{\cal R})^6}-
\frac{\alpha_{1}^i(0)\alpha_{2}^i(0)f_2(\bar n)}
{(2[\alpha_{1}^i(0)]^{1/3}+\Delta/{\cal R})^8}
\nn \\ &-&
\frac{\alpha_{1}^i(0)\alpha_{3}^i(0)f_3(\bar n)}
{(2[\alpha_{1}^i(0)]^{1/3}+\Delta/{\cal R})^{10}},
\end{eqnarray}
where ${\cal R}=[{\cal N}_f^{(1+\delta_1)}/{\cal N}_i]^{1/3}$.
This is the main result for the two-body vdW interaction. We can see from 
Eq.~(\ref{vdw2}) that the size-dependence of the two-body vdW interaction
depends upon the difference $\Delta$ between the true center-to-center
distance and the sum of the vdW radii. When $\Delta$ is close to 0, the 
vdW attraction reduces to the value it takes initially for 
$d=2[\alpha_{1}^i(0)]^{1/3}$ and we are almost 
unable to observe the size dependence of the vdW interaction. When
$\Delta$ is large and positive, the vdW attraction is rather weak. 
When $\Delta$ is negative, a stronger (e.g., covalent) bond may form 
between nanostructures. But in both cases, the size-dependence of the 
vdW interaction should be observed. We will demonstrate these 
observations with fullerenes, sodium clusters, and polycyclic aromatic 
hydrocarbons as follows.

\section{Three-body vdW interaction energy} 
In solids, a molecule or ion core not only interacts with another molecule 
or ion core, but also simultaneously 
interacts with other two or more species. Three-body vdW interaction is 
much weaker~\cite{Klein67,MBDoran71,OAL10,tao17} than two-body vdW 
interaction. Here we only consider the most important lowest-order 
dipole-dipole-dipole interaction. The asymptotic form of the three-body 
interaction energy is given by
$E_{\rm vdW}^{(3)}(\Iv\Jv\Kv)=C_9[3{\rm cos}(\theta_I)
{\rm cos}(\theta_J){\rm cos}(\theta_K)+1]/(d_{IJ}d_{IK}d_{JK})^3$,
where $d_{IJ}=|\Iv-\Jv|$ are the sides of a triangle, and $\theta_I$,
$\theta_J$, $\theta_K$ are the internal angles of the triangle formed by
$d_{IJ} d_{IK} d_{JK}$. $C_9$ is the three-body vdW coefficient, which can 
be calculated with the dipole polarizability from
\begin{eqnarray}\label{c9}
C_9=\frac{3}{\pi}\int_0^\infty du~\alpha_1^I(iu)
\alpha_1^J(iu)\alpha_1^K(iu). 
\end{eqnarray}
Substituting the dipole polarizability of Eq.~(\ref{sfapolar}) into the 
expression for the three-body vdW coefficient [Eq.~(\ref{c9})] and 
performing the integration over the imaginary frequency, we can obtain
an expression for the three-body coefficient. Next we consider 
identical nanostructures. Making use of an analysis similar to that 
for the two-body interaction, we can easily obtain the 
final expression for the size dependence of three-body vdW interaction 
energy
\begin{eqnarray}\label{three1}
E_{\rm vdW}^{(3)}
&=&
[\alpha_{1}^i(0)]^3f({\bar n},\theta_I,\theta_J,\theta_K) 
\{(2[\alpha_{1}^i(0)]^{1/3}+\Delta_{IJ}/{\cal R})
\nn \\ &\times&
(2[\alpha_{1}^i(0)]^{1/3}+\Delta_{IK}/{\cal R})
(2[\alpha_{1}^i(0)]^{1/3}+\Delta_{JK}/{\cal R})\}^{-3}.~~~~\nn
\end{eqnarray}
which is similar to the two-body vdW interaction, with ${\cal R}$ being 
defined below Eq.~(\ref{vdw2}). If the $\Delta_{IJ}$, $\Delta_{IK}$, and 
$\Delta_{JK}$ are all small, the three-body interaction is also size
independent. Otherwise, it is size-dependent. This is very similar to the
two-body interaction, but its effect on the total vdW interaction is 
small.

Summarizing, our size-dependence of the vdW interaction energy for 
identical nanostructures is formulated in terms of the hollow-sphere model 
within the SFA of Eq.~(\ref{sfapolar}). This model is valid for both
spherical and non-spherical nanostructures of slowly-varying densities, 
because non-sphericity~\cite{JTao14,Tao-Rappe16} can enter the model 
via the input static polarizability 
$\alpha_l(0)=(\alpha_{l,xx}+\alpha_{l,yy}+\alpha_{l,zz})/3$
obtained from accurate {\em ab initio} calculations. But 
Eq.~(\ref{vdw2}) is only instructive for quasi-spherical nanostructures 
at equilibrium center-to-center distance between them. It is least 
instructive for finite low-dimensional parallel nanostructures, for 
which the center-to-center distance at equilibrium is nearly independent 
of system size. In this case, the size-dependence of the vdW interaction 
energy mainly arises from the size dependence of vdW coefficients. As 
such, we can employ the asymptotic formula of Eq.~(\ref{vdw}) directly to 
study its size dependence, as examplified here with polycyclic aromatic 
hydrocarbons. For nanostructures of infinite length such as nanotubes, 
the situation is more complicated and will not be discussed. 

Finally, it is worth pointing out that the vdW interaction energy can 
also display a size-dependence in the case of nonzero $\Delta$, even when 
the vdW coefficients are additive. This suggests that this phenomena can be 
also described with an atom pairwise model~\cite{MScheffler09,SGrimme10,ERJ15}.

{\small
\begin{table*}
\caption{Terms of the van der Waals interaction energy between two identical 
polycyclic aromatic
hydrocarbons (PAHs) of AA stacking with the nearest neighbor center-to-center
separation $d=d_{cc}$~\cite{NatureM} obtained from the vdW-corrected vdW-DF-cx
nonlocal functional~\cite{KBerland14-1}. All quantities are in atomic units
(hartrees for energy, bohrs for distance). The vdW coefficients are evaluated
from the hollow-sphere model within the SFA of Eq.~(\ref{sfapolar}) with 
$t_l=R_l$, where the average valence electron density 
$\bar n = N/v$~\cite{Tao-Rappe16}. The reference values of $C_6$ are from 
TDDFT (time-dependent DFT) calculations~\cite{MALMarques07,AJiemchooroj05}. 
The input static dipole polarizabilities are taken from TDHF  
calculations~\cite{AJiemchooroj05}. The higher-order multipole
polarizabilities are estimated from the conventional formula of
Eq.~(\ref{conventionalP}). We have used a density ${\bar n} = 0.0468$.  
A minus sign ``$-$'' in front of all the vdW 
interactions has been suppressed.}
\begin{tabular}{l|cccccccccccccc}
\hline\hline
&\multicolumn{1}{c}{$\alpha_1$}&
\multicolumn{1}{c}{$C_6^{\rm ref}/10^2$}&
\multicolumn{1}{c}{$C_6/10^2$}&
\multicolumn{1}{c}{$C_8/10^4$}&
\multicolumn{1}{c}{$C_{10}/10^6$}&
\multicolumn{1}{c}{$d_{cc}$}&
\multicolumn{1}{c}{$\frac{d_{cc}}{d_{cc}(\rm C_6H_6\cdot C_6H_6)}$}&
\multicolumn{1}{c}{$(C_6/d^6) 10^2$}&
\multicolumn{1}{c}{$(C_8/d^8) 10^2$}&
\multicolumn{1}{c}{$(C_{10}/d^{10}) 10^2$}& \\ \hline
C$_6$H$_6$-C$_6$H$_6$              &68.23&17.73 &17.93 &15.65 &11.35&7.69&1.00 &0.87 &1.28 &1.57 \\
C$_{10}$H$_8$-C$_{10}$H$_8$        &117.3&48.67 &50.42 &63.16 &65.71&7.47&0.98 &2.90 &6.51 &12.1 \\
C$_{14}$H$_{10}$-C$_{14}$H$_{10}$  &176.6&100.3 &108.5 &178.5 &244.0&7.37&0.97 &6.77 &20.5 &51.6 \\
C$_{18}$H$_{12}$-C$_{18}$H$_{12}$  &244.8&175.1 &199.0 &407.1 &691.8&7.30&0.96 &13.1 &50.5 &161.0 \\
\hline\hline
\end{tabular}
\label{table2}
\end{table*}}

\section{Size-dependence study of nanostructures}
\subsection{Fullerenes} 
Fullerene is an important material with many remarkable 
properties such as great chemical stability and high sublimation or 
cohesive energy, leading to a variety of applications~\cite{RLoutfy02,
PJena10}. Its properties resemble those of graphene in the large-size 
limit, such as zero-energy gap and binding energy~\cite{NatureM}. 
In fullerene solids, the vdW interaction is dominantly important, while 
covalent interaction between fullerene molecules is negligibly 
small~\cite{TYR15}. The shape of fullerenes is quasispherical and the 
electron density on the surface of fullerenes is nearly uniform. 
Therefore, they are ideal model systems for the study of the vdW 
interaction energy. 

Table~\ref{table1} for fullerene solids shows that both the 
leading-order and higher-order vdW interactions, $-C_{2k}/d^{2k}$, 
are nearly size independent. This is because $\Delta$ values are all small, 
compared to the sum of vdW radii $d_{\rm vdW}=2[\alpha_1(0)]^{1/3}$. 
Table~\ref{table1} also shows the comparison of expensive TDHF calculations 
and model polarizability-based calculations for $C_6$. From 
Table~\ref{table1}, we can see the good agreement of our model $C_6$ 
with TDHF values. The size independence of fullerene pair interactions 
may not be valid for other near neighbor (NN) pair interactions, because 
for other NN pair interactions, $\Delta$ increases, while $d_{\rm vdW}$ is 
a constant. But the influence of other NN pair interactions is small, 
leading to the near size-independence of fullerene pair interaction. 
Table~\ref{table1} also shows the very slow convergence of the vdW series 
for the interaction energy between two fullerenes at equilibrium, as 
anticipated by Refs.~\cite{TYR15,PRSGC}.

The same analysis will also apply to the three-body vdW interaction energy. 
To demonstrate the size independence of the three-body interaction energy, 
we first evaluate the three-body vdW coefficient in each fullerene solid with 
Eq.~(\ref{c9}). Then we evaluate the three-body vdW interaction. The 
results are also displayed in Table~\ref{table1}. From Table~\ref{table1}, 
we can observe that, similar to the two-body vdW interaction, the three-body 
interaction is also size-independent, confirming our prediction.

\subsection{Sodium clusters}
Sodium cluster is a simple-metal cluster. It was shown that a sodium 
cluster can form a giant atom~\cite{SSaito87-1,SSaito87-2}. Therefore, the 
pair interaction between sodium clusters can form a covalent bond. This 
leads to an intermolecular distance between centers of two sodium clusters 
significantly shorter than the sum of the vdW radii of sodium clusters, 
$d_{\rm vdW}$. To confirm this, we have calculated the distance between 
the centers of mass of two identical sodium clusters $({\rm Na}_4)_2$ 
($D_{2h}$), $({\rm Na}_8)_2$ ($T_d$), and $({\rm Na}_{19})_2$ ($D_{5h}$) 
by putting them side by side, with Quantum ESPRESSO using the 
vdW-DF-cx functional~\cite{KBerland14-1}, which has proven accurate in 
nanostructures~\cite{NatureM}. In this calculation, the energy cutoff is 
30 hartree and only $\Gamma$ points are included in the k-mesh. The 
results are displayed in Table~\ref{table1}. From Table~\ref{table1}, we can see 
that cancellation of the size dependence between vdW coefficients and the vdW 
radii is incomplete. We have also repeated the calculation of the 
distance between the centers of mass of sodium cluster pair (Na$_8)_2$ by 
putting them head to head and head to tail (see Supplemental 
Material). The results are shorter, compared to that for side to 
side, suggesting that the cancellation of the size dependence between vdW 
coefficients and the vdW radii in incomplete in all the cases. In other 
words, the vdW interaction between sodium clusters is indeed size 
dependent. Table~\ref{table1} also shows the comparison of the vdW 
coefficients obtained from the {\em ab initio} method and the hollow-sphere 
model within the SFA. From Table~\ref{table1}, we 
see that the model calculation is in good agreement with more expensive 
{\em ab initio} values.

\begin{figure}
\includegraphics[angle=0.0,width=\columnwidth]{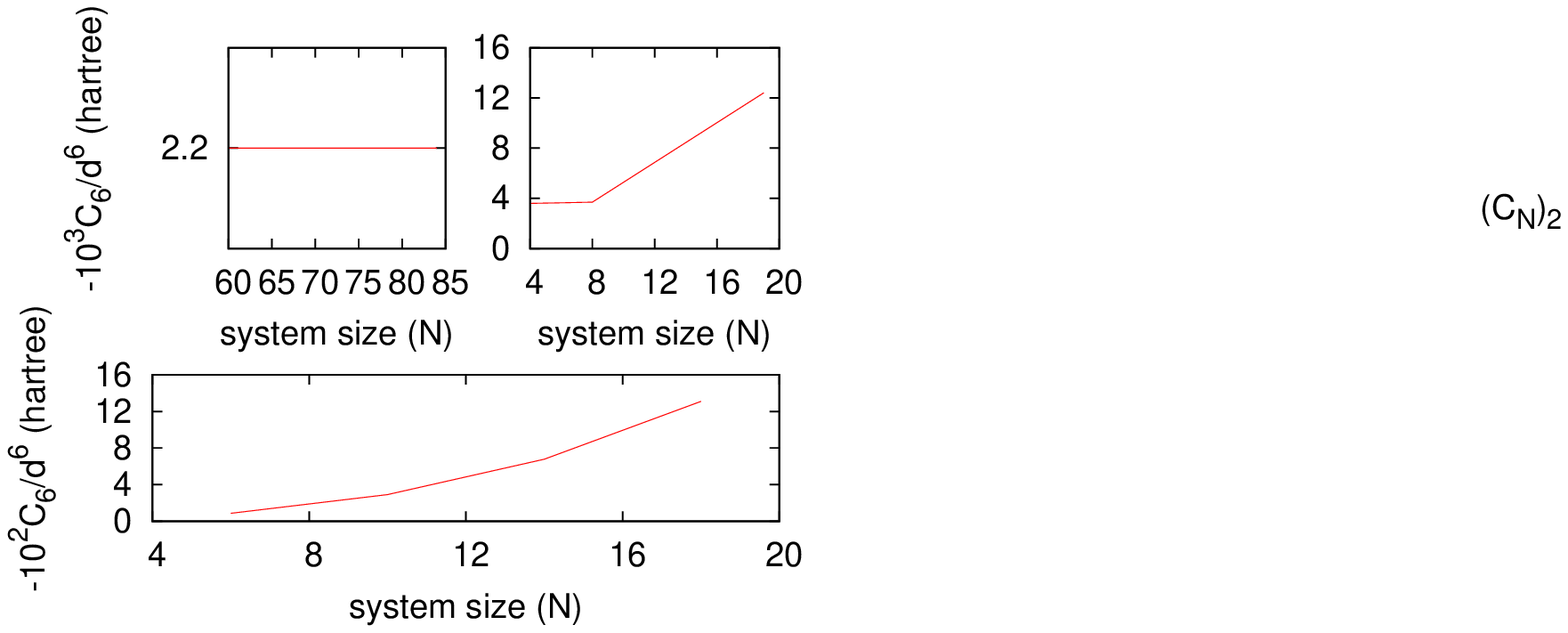}
\caption{Comparison of the size dependences for the absolute value of the 
vdW interaction energy term $-C_6/d^6$ between identical fullerenes in fcc solids 
(upper left panel), sodium clusters (upper right panel), and PAHs (lower panel).}
\label{figure1}
\end{figure}

\subsection{Polycyclic aromatic hydrocarbons (PAHs)} 
Finally, we have studied 
the size dependence of the vdW interaction energy between two identical 
PAHs with AA stacking. For such a geometry, a recent calculation with the 
vdW-DF-cx has shown~\cite{NatureM} that the 
plane-to-plane distance is only shrinking a little with system size, due 
to the fact that the vdW force also increases with system size, but the 
vdW coefficients per carbon atom increase rapidly with system size as we 
go from C$_6$H$_6$, C$_{10}$H$_8$, C$_{14}$H$_{10}$, to C$_{18}$H$_{12}$, 
owing to the nonadditivity. This leads to a rapid increase of the vdW 
interaction with system size. From Table~\ref{table2}, we can observe that, 
even when the molecules are highly non-spherical, the model vdW 
coefficients (Eq.~(\ref{sfapolar}) with $t_l=R_l$) are still accurate, 
compared to the more expensive time-dependent DFT-B3LYP calculations, 
suggesting the reliability of our results. In our calculations, the 
plane-to-plane distance of 2,3-Benzanthracene~\cite{NatureM} was used for 
C$_{18}$H$_{12}$-C$_{18}$H$_{12}$.

To see the damping effect on the vdW interaction energy, we calculate the
ratio of the center-to-center distance over the sum of the vdW radii 
$\frac{d_{cc}}{d_{\rm vdW}}$ and the center-to-center distance of PAHs over 
the benzene-to-benzene distance $\frac{d_{cc}}{d_{cc}(\rm C_6H_6\cdot C_6H_6)}$.
The former are all displayed in Table~\ref{table1}, while the latter are 
dispayed in Table~\ref{table2}. From Table~\ref{table1}, we see that 
all the ratios are nearly independent of system size, except for 
Na$_{19}$-Na$_{19}$. This means that the damping function should not change the 
size-dependence of the vdW interaction energy. The reason for this is 
the cancellation of the size-dependences of the equilibrium center-to-center 
distance and vdW radii. From Table~\ref{table2}, we can also see that the
ratios are nearly the same for PAHs, because the plane-to-plane distance of PAHs 
should not be size-dependent.  

Figure~\ref{figure1} shows the comparison of the dipole-dipole interaction 
energy term $-C_6/d^6$ for identical fullerene pairs, sodium cluster 
pairs, and PAH pairs. From Fig.~\ref{figure1}, we 
observe that the size dependence of fullerene pairs is nearly a constant, 
while that of PAH pairs yields the strongest size dependence. The 
size-dependence of the vdW interaction between sodium clusters is between 
these two extreme cases. From Tables~\ref{table1} and~\ref{table2}, we 
also see similar size dependences for higher-order interactions.

\section{Conclusion}
In conclusion, we employ the model polarizability and the experimental 
center-to-center distance (for fullerenes) or the distance from the vdW-DF-cx 
functional (for sodium clusters and PAHs) to study the size dependence of the vdW 
energy at the equilibrium distance. The former offers a good description of 
vdW coefficients via the Casimir-Polder formula, while the latter is accurate 
in the prediction of the center-to-center equibrium distance, but 
inaccurate in the vdW coefficients for some nanostructures such as 
fullerene~\cite{PRL12,NatureM}. The dependence of the vdW interaction energy
on system size or number of atoms is a common feature for nanostructures. 
It arises from the competition between the size-dependences of the vdW 
coefficients and the center-to-center distance. In this work, starting 
from the asymptotic long-range vdW interaction, we have derived an 
expression for the size-dependence of the vdW interaction, which is 
valid for identical nanostructures. We have studied the size-dependence 
of the vdW interaction for fullerenes, sodium clusters, and PAHs. Our 
calculations show that, for two identical nearest neighbor fullerenes 
in a fullerene solid, the vdW interaction is size-independent. This is 
unexpected, given that the vdW coefficients of fullerenes have very strong 
nonadditivity~\cite{JTao14} or non-linear effects, due to the electron 
delocalization. However, for low-dimensional nanostructures, the vdW 
interaction shows the strongest size-dependence. We illustrate this with 
planar PAHs. For sodium clusters, the size-dependence of the vdW 
interaction is between those of fullerenes and PAHs.

\section{Supplementary Material}
This material provides the details of different molecular geometries of 
sodium cluster. 

\section{Acknowledgements}
We thank Andrey Solov'yov for kindly sending us Ref. 65 with the atomic 
coordinates of sodium clusters. JT was supported by the NSF under Grant 
No. CHE 1640584. JT was also supported on Temple start-up from JPP. 
JPP and CS were supported by the NSF under Grant No. DMR-1607868. 
HT acknowledges support from the DOE Office of Science, Basic Energy 
Sciences (BES), under grant No. DE-SC0018194. Computational support 
was provided by HPC of Temple University and NERSC.


\begin{thebibliography}{100}
\bibitem{pbe96}
J.P. Perdew, K. Burke, and M. Ernzerhof, Phys. Rev. Lett. {\bf 77},
3865 (1996).

\bibitem{tpss03}
J. Tao, J.P. Perdew, V.N. Staroverov, and G.E. Scuseria,
Phys. Rev. Lett. {\bf 91}, 146401 (2003).

\bibitem{M06L}
Y. Zhao and D. G. Truhlar, J. Chem. Phys. {\bf 125}, 194101 (2006).

\bibitem{SCAN}
J. Sun, A. Ruzsinszky, and J.P. Perdew,
Phys. Rev. Lett {\bf 115}, 036402 (2015).

\bibitem{JSun16}
J. Sun, R.C. Remsing, Y. Zhang, Z. Sun, A. Ruzsinszky, H. Peng, Z. Yang, 
A. Paul, U. Waghmare, X. Wu, M.L. Klein, and J.P. Perdew,
Nat. Chem. {\bf 8}, 831 (2016).

\bibitem{TM16}
J. Tao and Y. Mo, Phys. Rev. Lett {\bf 117}, 073001 (2016).

\bibitem{excitation}
G. Tian, Y. Mo, and J. Tao,
J. Chem. Phys. {\bf 146}, 234102 (2017).

\bibitem{pbe0}
M. Ernzerhof and G.E. Scuseria, J. Chem. Phys. {\bf 110}, 5029 (1999).

\bibitem{Becke1996}
A. D. Becke, J. Chem. Phys. {\bf 104}, 1040 (1996).

\bibitem{peng16}
H. Peng, Z.-H. Yang, J. P. Perdew, and J. Sun,
Phys. Rev. X {\bf 6}, 041005 (2016).

\bibitem{tao-rappe14}
J. Tao and A.M. Rappe, 
Phys. Rev. Lett. {\bf 112}, 106101 (2014).

\bibitem{tao17}
J. Tao, F. Zheng, J. Gebhardt, J.P. Perdew, and A.M. Rappe,
Phys. Rev. Mater. {\bf 1}, 020802(R) (2017).   

\bibitem{Egger14}
D.A. Egger and L. Kronik,
J. Phys. Chem. Lett. {\bf 5}, 2728 (2014).

\bibitem{WAASaidi12}
W.A. Al-Saidi, V.K. Voora, and K.D. Jordan,
J. Chem. Theory Comput. {\bf 8}, 1503 (2012).

\bibitem{OHod12}
O. Hod,
J. Chem. Theory Compu. {\bf 8}, 1360 (2012).

\bibitem{ERJ12}
A. Otero-de-la-Roza and Erin R. Johnson,
J. Chem. Phys. {\bf 137}, 054103 (2012).

\bibitem{MScheffler12}
A. Tkatchenko, R.A. DiStasio, R. Car, and M. Scheffler,
Phys. Rev. Lett. {\bf 108}, 236402 (2012).

\bibitem{Grimme13}
T. Risthaus and S. Grimme,
J. Chem. Theory Compu. {\bf 9}, 1580 (2013).

\bibitem{book14}
J.G. Brandenburg and S. Grimme,
Top. Curr. Chem. {\bf 345}, 1 (2014).


\bibitem{ERJ15}
A. Otero-de-la-Roza and Erin R. Johnson,
J. Chem. Theory Comput. {\bf 11}, 4033 (2015).

\bibitem{TYR15}
J. Tao, J. Yang, and A.M. Rappe,
J. Chem. Phys. {\bf 142}, 164302 (2015).

\bibitem{BSachs11}
B. Sachs, T.O. Wehling, M.I. Katsnelson, and A.I. Lichtenstein, 
Phys. Rev. B {\bf 84}, 195414 (2011).

\bibitem{Anyan13}
T. Bu\u{c}ko, S. Leb\`{e}gue, J. Hafner, and J.G. \'{A}ngy\'{a}n,
Phys. Rev. B {\bf 87}, 064110 (2013).

\bibitem{TGould16}
T. Gould, S. Leb\`egue, J.G. \'Angy\'an, and T. Bu\'cko,
J. Chem. Theory Comput. {\bf 12}, 5920 (2016).

\bibitem{MScheffler09}
A. Tkatchenko and M. Scheffler,
Phys. Rev. Lett. {\bf 102}, 073005 (2009).


\bibitem{SGrimme10}
S. Grimme, J. Antony, S. Ehrlich, and H. Krieg,
J. Chem. Phys. {\bf 132}, 154104 (2010).


\bibitem{Becke07}
A.D. Becke and E.R. Johnson,
J. Chem. Phys. {\bf 127}, 154108 (2007).

\bibitem{Steinmann11-1}
S.N. Steinmann and C. Corminboeuf,
J. Chem. Phys. {\bf 134}, 044117 (2011).

\bibitem{Steinmann11-2}
S.N. Steinmann and C. Corminboeuf,
J. Chem. Theory Comput. {\bf 7}, 3567 (2011).


\bibitem{MDion04}
M. Dion, H. Rydberg, E. Schr\"oder, D.C. Langreth, and B.I. Lundqvist,
Phys. Rev. Lett. {\bf 92}, 246401 (2004).



\bibitem{KBerland14-1}
K. Berland and P. Hyldgaard,
Phys. Rev. B {\bf 89}, 035412 (2014).




\bibitem{review15}
K. Berland, V.R. Cooper, K. Lee, E. Schr\"oder,
T. Thonhauser, P. Hyldgaard, and B.I. Lundqvist,
Rep. Prog. Phys. {\bf 78}, 066501 (2015).

\bibitem{Voorhis09}
O.A. Vydrov and T. Van Voorhis,
Phys. Rev. Lett. {\bf 103}, 063004 (2009).


\bibitem{Granatier2011}
J. Granatier, P. Lazar, M. Otyepka, and P. Hobza,
J. Chem. Theory Comput. {\bf 7}, 3743 (2011).

\bibitem{JKlimes11}
J. Klimes, D.R. Bowler, and A. Michaelides,
Phys. Rev. B {\bf 83}, 195131 (2011).


\bibitem{PRL12}
A. Ruzsinszky, J.P. Perdew, J. Tao, G.I. Csonka, and J.M. Pitarke,
Phys. Rev. Lett. {\bf 109}, 233203 (2012).


\bibitem{DWR}
J.F. Dobson, A. White, and A. Rubio, 
Phys. Rev. Lett. {\bf 96}, 073201 (2006).

\bibitem{Klein67}
M.L. Klein and R.J. Munn, 
{\em J. Chem. Phys.} {\bf 47}, 1035 (1967).


\bibitem{OAL10}
O.A. von Lilienfeld and A. Tkatchenko,
J. Chem. Phys. {\bf 132}, 234109 (2010).

\bibitem{Bell66}
R.J. Bell and A.E. Kingston,
Proc. Phys. Soc. {\bf 88}, 901 (1966)

\bibitem{Tang69}
K.T. Tang,
Phys. Rev. {\bf 177}, 108 (1969).


\bibitem{MBDoran71}
M.B. Doran and I.J. Zucker,
J. Phys. C: Solid St. Phys. {\bf 4}, 307 (1971).

\bibitem{Paesani16}
F. Paesani, Acc. Chem. Res. {\bf 49}, 1844 (2016).


\bibitem{Chai00}
J.-D. Chai and M. Head-Gordon,
Phys Chem Chem Phys {\bf 10}, 6615 (2000).

\bibitem{KTTang84}
K.T. Tang and J.P. Toennies,
J. Chem. Phys. {\bf 80}, 3726 (1984).

\bibitem{wuyang02}
Q. Wu and W. Yang,
J. Chem. Phys. {\bf 116}, 515 (2002).

\bibitem{ERJ06}
E.R. Johnson and A.D. Becke,
J. Chem. Phys. {\bf 124}, 174104 (2006).

\bibitem{SGrimme11}
S. Grimme, S. Ehrlich, and L. Goerigk,
J. Comput. Chem. {\bf 32}, 1456 (2011).

\bibitem{SHPatil97}
S.H. Patil and K.T. Tang,
J. Chem. Phys. {\bf 106}, 2298 (1997).


\bibitem{JTao14}
J. Tao and J.P. Perdew, 
J. Chem. Phys. (Communication) {\bf 141}, 141101 (2014).

\bibitem{Tao-Rappe16}
J. Tao and A.M. Rappe, 
J. Chem. Phys. (Communication) {\bf 144}, 031102 (2016).

\bibitem{PRB16}
J. Tao, Y. Mo, G. Tian, and A. Ruzsinszky, 
Phys. Rev. B {\bf 94}, 085126 (2016).

\bibitem{GKG04}
G.K. Gueorguiev, J.M. Pacheco, and D. Tom\'anek, 
Phys. Rev. Lett. {\bf 92}, 215501 (2004).


\bibitem{NatureM}
J. Tao, Y. Jiao, Y. Mo, Z.-H. Yang, J.-X. Zhu, P. Hyldgaard, and J.P. Perdew, 
submitted.

\bibitem{JCP13}
J. Kauczor, P. Norman, W.A. Saidi,
J. Chem. Phys. {\bf 138}, 114107-1-8 (2013).

\bibitem{AJiemchooroj06}
A. Jiemchooroj, P. Norman, and B.E. Sernelius,
J. Chem. Phys. {\bf 125}, 124306 (2006).

\bibitem{RLoutfy02}
R. Loutfy and E. Wexler, 
Ablative and Flame-Retardant Properties of Fullerenes, in
{\em Perspectives of Fullerene Nanotechnology},
Part VII, 275-280. eds. by E. ${\bar {\rm O}}$sawa, 
(Kluwer Academic Publishers, New York, 2002).

\bibitem{PJena10}
{\em Nanoclusters, Volume 1: A Bridge across Disciplines},
edited by P. Jena and A.W. Castleman Jr.
(Elsevier, Amsterdam, 2010).

\bibitem{PRSGC} 
J.P. Perdew, A. Ruzsinszky, J. Sun. S. Glindmeyer, and G.I. Csonka, 
Phys. Rev. A {\bf 86}, 062714 (2012).

\bibitem{MNMagomedov05}
M.N. Magomedov,
High Temp. (USSR) {\bf 43}, 379 (2005).

\bibitem{RDBendale95}
R.D. Bendale and M.C. Zerner,
J. Phys. Chem. {\bf 99}, 13830 (1995).

\bibitem{DEBeck84}
D.E. Beck,
Phys. Rev. B {\bf 30}, 6935 (1984).

\bibitem{PRA02}
I.A. Solov'yov, A.V. Solov'yov†, and Walter Greiner,
Phys. Rev. A {\bf 65}, 053203 (2002).


\bibitem{QE}
P. Giannozzi {\em et al.},
J. Phys.:Condens.Matter {\bf 21}, 395502-1-19 (2009).



\bibitem{SSaito87-1}
S. Saito and S. Ohnishi,
Phys. Rev. Lett. {\bf 59}, 190 (1987). 


\bibitem{SSaito87-2}
Y. Ishii, S. Saito, and S. Ohnishi,
Z. Phys. D - Atoms, Molecules and Clusters {\bf 7}, 289 (1987).








\bibitem{MALMarques07}
M.A.L. Marques, A. Castro, G. Malloci, G. Mulas, and S. Botti,
J. Chem. Phys. {\bf 127}, 014107 (2007).


\bibitem{AJiemchooroj05}
A. Jiemchooroj, P. Norman, and B.E. Sernelius,
J. Chem. Phys. {\bf 123}, 124312 (2005).


\end{thebibliography}
\end{document}